\documentclass[12pt]{article}
\usepackage[utf8]{inputenc}
\usepackage{mathptmx}
\usepackage{geometry}
\usepackage{authblk}
\usepackage{amssymb}
\usepackage{amsmath}
\usepackage{amsthm}
\allowdisplaybreaks[4]
\usepackage{slashed}
\usepackage{graphicx,color,epsfig}
\usepackage{multirow}
\usepackage{cite}
\usepackage[colorlinks=true,linkcolor=red,citecolor=blue]{hyperref}
\begin{document}
\renewcommand{\arraystretch}{1.0}
\newcommand{\beq}{\begin{eqnarray}}
\newcommand{\eeq}{\end{eqnarray}}
\newcommand{\non}{\nonumber\\ }
\newcommand{\acp}{ {\cal A}_{CP} }
\newcommand{\psl}{ p \hspace{-1.8truemm}/ }
\newcommand{\nsl}{ n \hspace{-2.2truemm}/ }
\newcommand{\vsl}{ v \hspace{-2.2truemm}/ }
\newcommand{\epsl}{\epsilon \hspace{-1.8truemm}/\,  }
\title{Phenomenological analysis of the quasi-two-body $B \to D (R\to) K \pi$ decays in PQCD Approach}
\author[1]{Wen-Sheng Fang}
\author[1]{Zhi-Tian Zou}
\author[1]{Ying Li$\footnote{liying@ytu.edu.cn}$}
\affil[1]{\it \small Department of Physics, Yantai University, Yantai 264005, China}
\maketitle
\begin{abstract}
The quasi-two-body $B \to D (R\to) K \pi$ decays are calculated in PQCD approach based on the $k_T$ factorization by introducing the wave functions of $K\pi$ pair associated with the resonances $K^*(892)$, $K_0^*(1430)$ and $K_2^*(1430)$. The results show that most branching fractions are at the order of $10^{-7}$ or even smaller. However, for $B^0\to D^0(K^*\to)K\pi$ decays enhanced by the CKM element $V_{cs}$, their branching fractions are at the order of $10^{-6}$, which are measurable in the current ongoing experiments. Based on the narrow-width-approximation we also extract the branching fractions of the corresponding two-body $B \to D K^*$ decays and the results are in good agreement with previous predictions. Because these decays are only governed by the tree operators, there are no $CP$ asymmetries in these decays in standard model.
\end{abstract}

\section{Introduction}
It is known that rare $B$ decays play important roles in studying QCD, exploring the origin of $CP$ violation, and searching for possible effects of new physics beyond the standard model (SM). In past decades, the $B$ meson nonleptonic decays have received particular scrutiny at the $B$-factories Belle and BaBar, and the LHC-experiments. Besides two-body $B$ decay modes which have attached more attentions in past, many three-body $B$ decay modes have also been observed with branching fractions of order $10^{-5}$ in above experiments \cite{ParticleDataGroup:2022pth}. However, different from the two-body $B$ decays where the kinematics is fixed totally, the amplitude of a three-body decay depend on two invariant masses (e.g. $s_{12}$ and $s_{13}$ with the definition $m_{ij}\equiv(p_i+p_j)^2)$ . All the allowed physical kinematics define a triangle region in the $m_{12}-m_{13}$ plane and the density plot of the differential decay rate in this region is so-called Dalitz plot, which is widely used in analyzing the three-body decays in both experimental and theoretical studies. In addition, both resonant and nonresonant contributions are involved in the three-body decays, and the Dalitz plot technique enables us to analyze the different contributions. The Dalitz plot can be divided into different regions in term of the characteristic kinematics. The central region represents the nonresonant contributions, corresponding to the case that the three final particles fly apart with large energy $E\simeq m_B/3$. If one final particle is almost at rest and the other two particles fly back to back with energy $E\simeq m_B/2$, this case falls into the three corners of the Dalitz plot. Finally the edges of the Dalitz plot correspond to the situations that two of final particles move collinearly and the bachelor recoils back. In this case, two collinear particles might be produced from one intermediate resonance. Therefore, the study of the edges of the Dalitz plots enables us to probe the properties of the various resonances. In the theoretical side, many approaches have been proposed for analyzing the three-body $B$ meson decays, such as the QCD factorization  \cite{El-Bennich:2009gqk, Krankl:2015fha, Klein:2017xti, Cheng:2016shb, Cheng:2014uga, Li:2014oca}, the PQCD approach \cite{Wang:2016rlo, Li:2016tpn, Rui:2017bgg, Zou:2020atb, Zou:2020fax, Zou:2020ool}, and methods based on symmetries \cite{Zhang:2013oqa, El-Bennich:2006rcn, Hu:2022eql}.

Specifically, three-body $B$ decays have been used to extract the CKM matrix weak angles $\alpha$, $\beta$ and $\gamma$. For instance, the charmed three-body $B^0\to D K^+\pi^-$ decay can used to constrain the CKM angle $\gamma$ within the interference between $\bar{b}\to \bar{c}u\bar{s}$ and $\bar{b}\to\bar{u}c\bar{s}$ amplitudes. For the decays $B^0\to \overline {D^0}(K^{*0}\to) K^+\pi^-$ and $B^0\to D^0 (K^{*0}\to) K^+\pi^-$ that are induced by $\bar{b}\to \bar{c}u\bar{s}$ and $\bar{b}\to\bar{u}c\bar{s}$ respectively, the two  amplitudes are close in magnitude, leading to sizable direct $CP$ asymmetries in decays $B\to D_\pm  K^+\pi^-$. Moreover, $B^0\to D K^+\pi^-$ decays are especially advantageous since the charge of the kaon unambiguously tags the flavor of the decaying $B$ meson, obviating the need for time-dependent analysis. This appears to be one of the most promising channels to make a precise measurement of $\gamma$ \cite{Atwood:2000ck, Atwood:1996ci}. In recent years, LHCb collaboration has performed the Dalitz plot analysis and measured the branching fractions of the $B^0\to \overline{D}^0\pi^+\pi^-$, $B^0\to D(\overline{D}^0) K^+\pi^-$, $B^0\to \overline{D}^{*0}K^+\pi^-$, $B_s^0\to \overline{D}^{*0}K^-\pi^+$, and $B^+\to D^-K^+\pi^+$ decays \cite{LHCb:2015klp, LHCb:2015tsv, LHCb:2016bsl, LHCb:2021jfy, LHCb:2015eqv} for improving the precision to the CKM angles of existing and studying the properties of various resonances. Motivated by the released results of LHCb, we have investigated the CKM-favoured charmed three-body $B_{(s)}\to \overline{D}K\pi$ decays with $P$-wave resonances $K^*(892)$ and $K^*(1410)$, the $S$-wave resonance $K_0^*(1430)$ and the $D$-wave resonance $K_2^*(1430)$ in PQCD approach by adopting an appropriate two-meson wave functions of $K\pi$ pair \cite{Zou:2022xrr}. The theoretical results agree well with the experimental data. To keep completeness, we shall extend our studies to the $B\to D K\pi$ decays also with the $K^*(892)/(1410)$, $K_{0/2}^*(1430)$ resonances using the well-determined wave functions of the $K\pi$ pair.

The  paper is organized as follows. In Sect.~\ref{sect:2}, we shortly review the formalism of PQCD approach in association with the wave functions of meson and $K\pi$-pair, and then present the perturbative calculations of considered decays. The decay amplitudes are also collected in this section. The numerical results and some discussions are given in Sect.~\ref{sect:3}. Finally,we summarize this work is in Sect.~\ref{sect:4}.

\section{Decay Formalism and Decay Amplitudes}\label{sect:2}
For clarity, we define the kinematics of the three-body decay as
\begin{eqnarray}
B(p_B)\to M_1(p_1)+M_2(p_2)+M_3(p_3),
\end{eqnarray}
and it is customary to take these variables as two invariant masses of two pairs of final state particles,
\begin{eqnarray}
m_{12}^2=(p_1+p_2)^2, m_{13}^2=(p_1+p_3)^2.
\end{eqnarray}
Thus, the amplitude of the three-body decay is a function of the two kinematic variables, $m_{12}^2$ and $m_{13}^2$. As aforementioned, the central parts of three edges mean that two particles move collinearly with large energy and the other particle recoils back. In this case, the interactions between the meson-pair and the bachelor particle are power suppressed naturally. This kind of process is also called quasi-two-body process. The interactions in the meson-pair can be absorbed into a two-meson wave function. Thus, the quasi-two-body decay is very similar to a two-body decay, and the factorization formula would be applied by replacing one final particle by the meson-pair.

It should be stressed that the theoretical description of the three-body $B$ decays is still in the stage of modeling, and the isobar model \cite{Sternheimer:1961zz,Herndon:1973yn} and the K-matrix formalism \cite{Chung:1995dx} are usually applied in the Dalitz plot analysis of experimental data, especially the isobar model. Based on the isobar model,the decay amplitude can be decomposed into a coherent sum of amplitudes from $N$ individual decay channels with different resonances,
\begin{eqnarray}
\mathcal{A}=\sum_{i=1}^{N}a_i\mathcal{A}_i,
\end{eqnarray}
where $\mathcal{A}_i$ is the amplitude of one quasi-two-body decay with respect to a certain resonance $R_i$. The complex coefficient $a_j$ reflecting the relevant magnitude and the relative phase of the different channels can be determined from the experimental data, while the amplitude $\mathcal{A}_i$ can be theoretically calculated within QCD-inspired approach. In this work, the PQCD approach that is based on the $k_T$ factorization will be employed, where the spectator quark is kicked by a hard gluon.

In the framework of PQCD, the amplitude of a charmed quasi-two-body $B$ meson decay can be written as a convolution
\begin{eqnarray}
\mathcal{A}_i=\Phi_B\otimes\mathrm{H}\otimes J \otimes S \otimes\Phi_{M_1M_2,i}\otimes\Phi_D,
\label{amp}
\end{eqnarray}
all of which are well-defined and gauge-invariant. $J$ and $S$ denote the jet function from threshold resummation and the Sudakov factor from $k_T$ resummation, respectively. $\Phi_B$ and $\Phi_D$ are the wave functions of the $B$ meson and $D$ meson, describing how two inner quarks are combined into the heavy mesons. $\Phi_{M_1M_2,i}$ is the wave function of $K\pi$ pair, where both the soft interactions between two mesons and the contributions from $R_i$ are included. The hard kernel $\rm H$ for the $b$ quark decay, similar to the two-body case, starts with the diagrams of single hard gluon exchange, and can be calculated perturbatively.

The relevant effective weak Hamiltonian of $\bar{b}\to \bar{u} c\bar{q}~(q=s,d)$ decay is give by \cite{Buchalla:1995vs}
\begin{eqnarray}
\mathcal{H}_{eff}=\frac{G_F}{\sqrt{2}}V_{ub}^*V_{cq}\Big[C_1O_1+C_2O_2\Big],
\end{eqnarray}
where the $V_{ub}$ and $V_{cq}$ are the CKM matrix elements. $C_1$ and $C_2$ are the wilson coefficients (WCs) corresponding to the tree level four-quark current-current operators $O_1$ and $O_2$, respectively. The tree operators $O_{1,2}$ are given as
\begin{eqnarray}
O_1&= \bar{b}_{\alpha}\gamma^{\mu}(1-\gamma_5)u_{\beta} \bar{c}_{\beta}\gamma_{\mu}(1-\gamma_5)q_{\alpha} ,\nonumber\\
O_2&= \bar{b}_{\alpha}\gamma^{\mu}(1-\gamma_5)u_{\alpha} \bar{c}_{\beta}\gamma_{\mu}(1-\gamma_5)q_{\beta},
\end{eqnarray}
where $\alpha$ and $\beta$ are the color indexes.

\begin{figure}[htb]
\begin{center}
\centering\includegraphics [scale=0.5] {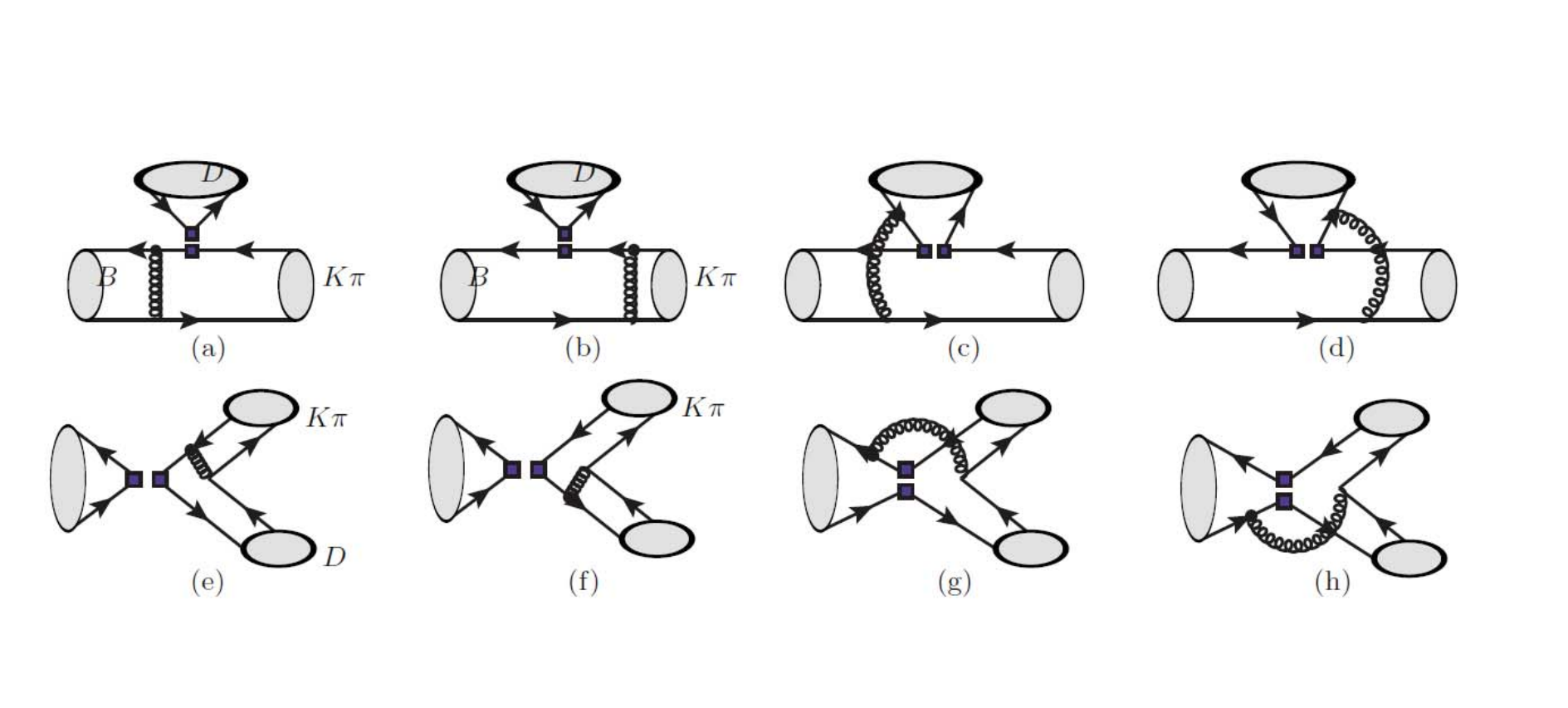}
\caption{Leading quark-level Feynman diagrams for for the $B \to D (R\to) K \pi$ decays.}  \label{fig:fig1}
\end{center}
\end{figure}

The six-quark hard kernel $\rm H$ consists of the diagrams with at least one hard gluon. The complete set of leading-order diagrams for the $B \to D (R\to) K \pi$ decays is displayed in Fig.~\ref{fig:fig1}. Figures.~\ref{fig:fig1}(a) and \ref{fig:fig1}(b), referred to as the factorizable emission diagrams, correspond to the leading contribution in QCDF. Figures~\ref{fig:fig1}(c) and  \ref{fig:fig1}(d), are referred to as the nonfactorizable emission diagrams. Figures~\ref{fig:fig1}(e) and ~\ref{fig:fig1}(f), and Figures.~\ref{fig:fig1}(g) and ~\ref{fig:fig1}(h) are referred to as the factorizable annihilation diagrams and the nonfactorizable annihilation ones, respectively.

We note that the wave functions of the $B$ meson and $D$ meson have been well defined in the two-body $B$ decays. In contrast, the wave functions of $K\pi$ pair with respect to the $S$-wave, $P$-wave, and $D$-wave $K^*$ resonances are less studied. The $S$-wave $K\pi$ pair wave function $\Phi_{S,K\pi}$ is given as \cite{Li:2019hnt,Wang:2020saq},
\begin{eqnarray}
\Phi_{S,K\pi}=\frac{1}{2\sqrt{N_c}}\left[P\mkern-10.5mu/\phi_S^a(z,\zeta,\omega)+\omega\phi_S^s(z,\zeta,\omega)+\omega(n\mkern-10.5mu/ v\mkern-10.5mu/-1)\phi_S^t(z,\zeta,\omega)\right],
\end{eqnarray}
where $P$ and $\omega$ are the momentum and the invariant mass of the $K\pi$ pair respectively, satisfying $P^2=\omega^2$. The dimensionless vectors $n=(1,0,\mathbf{0}_T)$ and $v=(0,1,\mathbf{0}_T)$ are the light-like vectors. $\phi_S^a$ and $\phi_S^{s,t}$ are the twist-2 and twist-3 distribution amplitudes (DAs), respectively. The inner parameter $z$ is the momentum fraction of the spectator quark, and $\xi$ is the momentum fraction of the $K$ meson in the $K\pi$ pair. For the wave function of the $P$-wave $K\pi$ pair, due to the law of conservation of angular momentum, we here only consider the longitudinal polarization component, which is given as
\begin{eqnarray}
\Phi_{P,K\pi}=\frac{1}{\sqrt{2N_C}}\left[P\mkern-10.5mu/\phi_P^a(z,\xi,\omega)+\omega\phi_P^s(z,\xi,\omega)
+\frac{P\mkern-10.5mu/_1P\mkern-10.5mu/_2-P\mkern-10.5mu/_2P\mkern-10.5mu/_1}{\omega(2\xi-1)}\phi_P^t(z,\xi,\omega)\right],
\label{pwave}
\end{eqnarray}
where $\phi_P^a$ is the twist-2 DA, and $\phi_P^{s,t}$ are twist-3 ones. For the same reason, the behavior of the $D$-wave $K\pi$ pair is very similar to the $P$-wave one \cite{Cheng:2010hn}, and only differences are the DAs $\phi_D^a$, $\phi_D^s$, and $\phi_D^t$. All distribution amplitudes have been determined from the experimental data, and readers are referred to our previous study \cite{Zou:2022xrr}. It should be emphasized that the DA contains both resonant and non-resonant contributions. Different from the DA of meson, the time-like form factor is introduced to describe the resonant contribution.

Based on the factorization formula and the Hamiltonian introduced above, we could calculate the decay amplitudes. Because both $O_1$ and $O_2$ are $(V-A)(V-A)$ current, there are only four kinds of amplitudes marked as $\mathcal{F}$, $\mathcal{M}$, $\mathcal{A}$  and $\mathcal{W}$, corresponding to the factorizable emission diagrams, nonfactorizable emission diagrams, the factorizable annihilation diagrams and the nonfactorizable annihilation diagrams, respectively. We now calculate the four kinds of amplitudes corresponding to the diagrams with different resonances. At first, the decay amplitudes including $S$-wave $K\pi$ pair are listed as
{\small
\begin{eqnarray}
\mathcal{F}&=&8\pi C_F m_B^4f_D\int_0^1dx_1dx_3\int_0^{\infty}b_1db_1b_3db_3\phi_B(x_1,b_1)\nonumber\\
&&\times\Bigg\{\Bigg[\Big(1+x_2-(1+2x_2)r_2^2-(1+x_2)r_3^2\Big)\phi_{X}^a(x_3)\nonumber\\
&&+(1-3x_3)r_3\phi_{S}^t(x_3)+(1-2x_3)r_3\phi_{X}^s(x_3)\Bigg]E_{ef}(t_a)h_{ef}[x_1,x_3(1-r_2^2),b_1,b_3]\nonumber\\
&&-\Bigg[r_3^2\phi_{S}^a(x_3)-2r_3\phi_{X}^s(x_3)\Bigg]E_{ef}(t_b)h_{ef}[x_3,x_1(1-r_2^2),b_3,b_1]\ \Bigg\},
\\
\mathcal{M}&=&16\sqrt{\frac{2}{3}}C_F\pi
m_B^4\int_0^1dx_1dx_2dx_3\int_0^{\infty}b_1db_1b_2db_2\phi_B(x_1,b_1)\phi_{D}(x_2)\nonumber\\
&&\times\Bigg\{\Bigg[\Big((x_3-2x_2)r_3^2+x_2\Big)\phi_{X}^a(x_3)-r_3x_3\phi_{X}^s(x_3)+r_3x_3\phi_{X}^t(x_3)\Bigg]
E_{enf}(t_c)h_{enf}(\alpha,\beta_1,b_1,b_2)\nonumber\\
&&+\Bigg[\Big(1+x_3-x_2-(2x_3+1-x_2)r_2^2-(x_3+2-2x_2)r_3^2\Big)\phi_{X}^a(x_3)-x_3r_3\phi_{X}^t(x_3)\nonumber\\
&&-x_3r_3\phi_{X}^s(x_3)\Bigg]E_{enf}(t_d)h_{enf}(\alpha,\beta_2,b_1,b_2)\Bigg\},
\\
\mathcal{A}&=&-8C_F f_B\pi m_B^4\int_0^1dx_2dx_3\int_0^{\infty}b_2db_2b_3db_3\phi_{D}(x_3)\nonumber\\
&&\times\Bigg\{\Bigg[\Big((1-2x_3)r_2^2+x_3-x_3r_3^2\Big)\phi_{X}^a(x_2) +2(1+x_3)r_2r_3\phi_{X}^s(x_2)\Bigg]E_{af}(t_e)h_{af}(\alpha_1,\beta,b_2,b_3)\nonumber\\
&&+\Bigg[\Big((1+2x_3)r_3^2-x_2+x_2r_2^2\Big)\phi_{X}^a(x_2)+(1-2x_3)r_2r_3\phi_{X}^t(x_2)\nonumber\\
&&-(1+2x_3)r_2r_3\phi_{X}^s(x_2)\Bigg]E_{af}(t_f)h_{af}(\alpha_2,\beta,b_2,b_3)\ \Bigg\},
\\
\mathcal{W}&=&16\sqrt{\frac{2}{3}}C_F \pi m_B^4\int_0^1dx_1dx_2dx_3\int_0^{\infty}b_1db_1b_2db_2 \phi_B(x_1,b_1)\phi_{D}(x_3)\nonumber\\
&&\times\Bigg\{\Bigg[\Big(r_2^2+(1-x_3+2x_2)r_3^2-x_2\Big)\phi_{X}^a(x_2)-r_2r_3\Big((2+x_3+x_2)\phi_{X}^s(x_2)\nonumber\\
&&+(x_2-x_3)\phi_{X}^t(x_2)\Big)\Bigg]E_{anf}(t_g)h_{anf}(\alpha,\beta_1,b_1,b_2)\nonumber\\
&&+\Bigg[\Big(x_3+(x_2-2x_3)r_2^2\Big)\phi_{X}^a(x_2)+r_2r_3\Big((x_3+x_2)\phi_{X}^s(x_2)\nonumber\\
&&-(x_2-x_3)\phi_{X}^t(x_2)\Big)\Bigg]E_{anf}(t_h)h_{anf}(\alpha,\beta_2,b_1,b_2)\ \Bigg\},
\end{eqnarray}
}
with $X$ denoting $S$,$P$,$D$ for different waves. The hard functions $h_{ef, enf, af, anf}$, the dynamic scales $t_i$, the Sudakov form factors $E_{ef, enf, af, anf}$ and the distribution amplitudes $\phi_B$, $\phi_D$ and $\phi_X^{a,s,t}$ in the wave functions of mesons and meson-pairs are referred to \cite{Zou:2012sx}.

Within the analytic amplitudes of the each diagrams contributing to the quasi-two-body $B_{(s)}\to {D}K\pi$ decays at the leading order in PQCD approach, we then obtain the total decay amplitudes with the CKM matrix elements and the WCs, which are presented as
\begin{eqnarray}
&&\mathcal{A}(B^0 \to \ D^0(K^+\pi^-))=\frac{G_F}{\sqrt{2}}V_{ub}^*V_{cs}\Bigg[\left(C_1+\frac{C_2}{3}\right)\mathcal{F}+C_2\mathcal{M}\Bigg],\\
&&\mathcal{A}(B_s \to \
D^{0}(K^- \pi^+))=\frac{G_F}{\sqrt{2}}V_{ub}^*V_{cd}\Bigg[\left(C_1+\frac{C_2}{3}\right)\mathcal{F} +C_2\mathcal{M} \Bigg],\\
&&\mathcal{A}(B_s \to \ D^+(K^- \pi^0))=\frac{G_F}{\sqrt{2}}V_{ub}^*V_{cd}\Bigg[\left(\frac{C_1}{3}+C_2\right)\mathcal{F} +C_1\mathcal{M} \Bigg],\\
&&\mathcal{A}(B^+ \to \ D^{0}(K^+ \pi^0))=\frac{G_F}{\sqrt{2}}V_{ub}^*V_{cs}\Bigg[\left(C_1+\frac{C_2}{3}\right)\mathcal{F}+C_2\mathcal{M} +\left(\frac{C_1}{3}+C_2\right)\mathcal{A} +C_1\mathcal{W} \Bigg],\\
&&\mathcal{A}(B_s \to D_s^+(K^- \pi^0))=\frac{G_F}{\sqrt{2}}V_{ub}^*V_{cs}\Bigg[\left(\frac{C_1}{3}+C_2\right)\mathcal{F} +C_1\mathcal{M}
+\left(C_1+\frac{C_2}{3}\right)\mathcal{A} +C_2\mathcal{W} \Bigg],\\
&&\mathcal{A}(B^0 \to D_s^+(K^- \pi^0))=\frac{G_F}{\sqrt{2}}V_{ub}^*V_{cd}\Bigg[\left(C_1+\frac{C_2}{3}\right)\mathcal{A} +C_2\mathcal{W}\Bigg],\\
&&\mathcal{A}(B^+ \to \ D^+( K^+ \pi^-))=\frac{G_F}{\sqrt{2}}V_{ub}^*V_{cs}\Bigg[\left(\frac{C_1}{3}+C_2\right)\mathcal{A} +C_1\mathcal{W} \Bigg],\\
&&\mathcal{A}(B^+ \to D_s^+(K^- \pi^+))=\frac{G_F}{\sqrt{2}}V_{ub}^*V_{cd}\Bigg[\left(\frac{C_1}{3}+C_2\right)\mathcal{A} +C_1\mathcal{W} \Bigg].
\end{eqnarray}

For the nonleptonic charmless two-body $B$ decays, the amplitudes of two nonfactorizable emission diagrams are cancelled by each other due to negligible masses of light quarks. However, for the charmed $B$ decays with the $D$ meson emission, two amplitudes no longer cancel each other, as the mass of the charm quark is much larger than that of light quark. In addition, though the amplitudes of this kind of decays are suppressed by the CKM elements, they are enhanced by the large WC $C_2$. So, one expects that the  branching fractions of these CKM suppressed decays are comparable with those of CKM-favored but color-suppressed decays.

\section{Numerical Results and Discussions}\label{sect:3}
In this section, we firstly present the parameters used in our numerical calculations, including the QCD scale, the masses and the lifetimes of the $B$ mesons, the mass of the $D$ meson, the masses and the widths of the intermediate resonances, and the CKM matrix elements,
\begin{eqnarray}
&&\Lambda_{QCD}^{f=4}=0.25\pm0.05~{\rm GeV},\;m_{B}=5.279~ {\rm GeV},\;m_{B_s}=5.366~ {\rm GeV},\nonumber\\
&&\tau_{B^+}=1.638~ {\rm ps},\;\tau_{B^0}=1.519 ~ {\rm ps},\;\tau_{B_s}=1.520~ {\rm ps},\nonumber\\
&&m_{D^0}=1.865~ {\rm GeV},\;m_{D^+}=1.869~{\rm GeV},\;m_{D_s}=1.968~{\rm GeV},\nonumber\\
&&m_{K^*(892)}=0.892 ~{\rm GeV},\;m_{K^*(1410)}=1.414~ {\rm GeV},\;m_{K_0^*(1430)}=1.425~ {\rm GeV},\nonumber\\
&&m_{K^*_2(1430)}=1.427~{\rm GeV},\;\Gamma_{K^*(892)}=51.4 ~{\rm GeV},\;\Gamma_{K^*(1410)}=232~ {\rm MeV},\nonumber\\
&&\Gamma_{K^*_0(1430)}=270~ {\rm MeV},\;\Gamma_{K_2^*(1430)}=100~ {\rm MeV},\nonumber\\
&&V_{cd}=0.22486,\;V_{ub}=0.00369\pm0.00011,\;V_{cs}=0.97349.
\end{eqnarray}

\begin{table}[!t]
\caption{The branching ratio (in $10^{-6}$) of  the three-body $B/B_s\to DK\pi$ decays with the resonance $K_0^*(1430)$ in the PQCD appraoch}
\begin{center}
\begin{tabular}{l c c}
 \hline \hline
 \multicolumn{1}{c}{Decay Modes}&\multicolumn{1}{c}{Class}&\multicolumn{1}{c}{PQCD }  \\
\hline\hline

$B^0 \to \ D^0K^+\pi^-(LASS)$&
&$1.71^{+0.83+0.46+0.12}_{-0.5-0.39-0.05}$    \\

$B^0 \to \ D^0(K_0^{*0}(1430)\to) K^+\pi^-$
& C
&$1.21^{+0.56+0.31+0.05}_{-0.49-0.33-0.09}$   \\

$B^0 \to \ D^0K^+\pi^- (LASS NR)$ &
& $0.97^{+0.39+0.24+0.04}_{-0.36-0.24-0.07}$    \\

\hline

$B_s \to \ D^{0} K^-\pi^+(LASS)$  &
&$(6.50^{+3.60+0.90+0.20}_{-1.80-1.10-0.40})\times 10^{-2}$    \\

$B_s \to \ D^{0}(\overline{K}_0^{*0}(1430)\to) K^- \pi^+$
&C
&$(4.30^{+2.90+0.80+0.30}_{-1.20-0.80-0.20})\times 10^{-2}$   \\

$B_s \to \ D^{0}K^-\pi^+(LASS NR)$
& & $(3.50^{+1.90+0.50+0.10}_{-1.00-0.70-0.30})\times 10^{-2}$    \\

\hline

$B^+ \to \ D^{0}K^+\pi^0(LASS)$      &
&$0.67^{+0.15+0.30+0.00}_{-0.45-0.30-0.15}$   \\

$B^+ \to \ D^{0}(K_0^{*+}(1430)\to) K^+ \pi^0$
&C
&$0.39^{+0.16+0.23+0.06}_{-0.21-0.16-0.08}$  \\

$B^+ \to \ D^{0}K^+\pi^0(LASS NR)$   &
& $0.34^{+0.08+0.17+0.00}_{-0.20-0.14-0.08}$    \\

\hline

$B_s \to \ D^+K^-\pi^0(LASS)$       &
&$0.33^{+0.19+0.02+0.02}_{-0.07-0.02-0.02}$    \\

$B_s \to \ D^+(K^{*-}(1430)\to) K^- \pi^0$
&T
&$0.29^{+0.17+0.02+0.02}_{-0.13-0.02-0.02}$   \\

$B_s \to \ D^+K^-\pi^0(LASS NR)$  &
& $0.18^{+0.10+0.01+0.01}_{-0.04-0.01-0.01}$    \\
\hline
$B_s \to D_s^+K^-\pi^0(LASS)$       &
&$7.80^{+4.07+0.37+0.46}_{-1.73-0.55-0.48}$   \\

$B_s \to D_s^+(K_0^{*-}(1430)\to) K^+ \pi^0$
& T
&$7.10^{+3.70+0.37+0.42}_{-1.55-0.48-0.41}$   \\

$B_s \to D_s^+K^-\pi^0(LASS NR)$  &
& $4.43^{+2.32+0.22+0.25}_{-0.98-0.31-0.27}$     \\

\hline

$B^0 \to D_s^+K^-\pi^0(LASS)$         &
&$(2.00^{+0.35+0.01+0.15}_{-0.45-0.20-0.15})\times 10^{-2}$    \\

$B^0 \to D_s^+(K_0^{*-}(1430)\to) K^- \pi^0$
&E
&$(1.40^{+0.45+0.10+0.20}_{-0.28-0.10-0.03})\times 10^{-2}$   \\

$B^0 \to D_s^+K^-\pi^0(LASS NR)$   &
& $(1.00^{+0.2+0.02+0.10}_{-0.30-0.10-0.10})\times 10^{-2}$     \\
\hline
$B^+ \to \ D^+K^+\pi^-(LASS)$       &
&$0.11^{+0.08+0.04+0.04}_{-0.09-0.02-0.00}$    \\

$B^+ \to \ D^+(K_0^{*0}(1430)\to) K^+ \pi^-$
&A
&$0.09^{+0.08+0.04+0.04}_{-0.09-0.02-0.00}$   \\

$B^+ \to \ D^+K^+\pi^-(LASS NR)$  &
& $0.06^{+0.07+0.02+0.04}_{-0.03-0.02-0.00}$     \\
 \hline
$B^+ \to D_s^+K^-\pi^+(LASS)$      &
&$(0.90^{+0.60+0.50+0.40}_{-0.40-0.30-0.00})\times 10^{-2}$   \\

$B^+ \to D_s^+(K_0^{*0}(1430)\to) K^- \pi^+$
&A
&$(0.76^{+0.50+0.20+0.20}_{-0.40-0.20-0.00})\times 10^{-2}$  \\

$B^+ \to D_s^+K^-\pi^+(LASS NR)$   &
& $(0.50^{+0.30+0.10+0.20}_{-0.40-0.20-0.00})\times 10^{-2}$    \\

 \hline \hline
\end{tabular}\label{tb1}
\end{center}
\end{table}

\begin{table}[!t]
\caption{The branching ratios (in $10^{-6}$) of the three-body  $B/B_s \to D K\pi$ decays with the resonances $K^*(892)$ and $K^*(1410)$ in the PQCD approach.}
\begin{center}
\begin{tabular}{l c c}
 \hline \hline
 \multicolumn{1}{c}{Decay Modes}&\multicolumn{1}{c}{Class }&\multicolumn{1}{c}{PQCD }  \\
\hline\hline
$B^0 \to \ D^0(K^{*0}(892)\to) K^+\pi^-$
& C
&$1.96^{+1.01+0.52+0.11}_{-0.87-0.41-0.12}$  \\

$B^0 \to \ D^0(K^{*0}(1410)\to) K^+\pi^-$
&C
&$0.21^{+0.09+0.06+0.01}_{-0.08-0.05-0.01}$  \\

\hline
$B_s \to \ D^{0}(\overline{K}^{*0}(892)\to) K^- \pi^+$
&C
&$0.08^{+0.05+0.02+0.00}_{-0.03-0.02-0.00}$ \\

$B_s \to \ D^{0}(\overline{K}^{*0}(1410)\to) K^- \pi^+$
&C
&($0.87^{+0.51+0.2+0.05}_{-0.27-0.2-0.05})\times 10^{-2}$ \\
\hline
$B^+ \to \ D^{0}(K^{*+}(892)\to) K^+ \pi^0$
& C
&$1.00^{+0.43+0.20+0.00}_{-0.48-0.27-0.07}$  \\

$B^+ \to \ D^{0}(K^{*+}(1410)\to) K^+ \pi^0$
&C
&$0.09^{+0.05+0.03+0.02}_{-0.03-0.01-0.00}$  \\
\hline
$B_s \to \ D^+(K^{*-}(892)\to) K^- \pi^0$
&T
&$0.6^{+0.30+0.03+0.04}_{-0.15-0.04-0.04}$  \\

$B_s \to \ D^+(K^{*-}(1410)\to) K^- \pi^0$
&T
&($7.00^{+3.00+0.30+0.40}_{-2.00-0.40-0.40})\times 10^{-2}$ \\
\hline
$B_s \to D_s^+(K^{*-}(892)\to) K^- \pi^0$
&T
&$13.3^{+6.84+0.76+0.80}_{-3.04-0.73-0.79}$  \\

$B_s \to D_s^+(K^{*-}(1410)\to) K^- \pi^0$
&T
&$1.50^{+0.77+0.08+0.09}_{-0.33-0.08-0.09}$  \\
\hline
$B^0 \to D_s^+(K^{*-}(892)\to) K^- \pi^0$
&E
&$(0.53^{+0.30+0.10+0.00}_{-0.30-0.10-0.03})\times 10^{-2}$  \\

$B^0 \to D_s^+(K^{*-}(1410)\to) K^- \pi^0$
&E
&$(0.05^{+0.03+0.01+0.00}_{-0.03-0.02-0.01})\times 10^{-2}$  \\
\hline
$B^+ \to \ D^+(K^{*0}(892)\to) K^+ \pi^-$
&A
&$0.21^{+0.10+0.03+0.04}_{-0.06-0.02-0.00}$  \\

$B^+ \to \ D^+(K^{*0}(1410)\to) K^+ \pi^-$
&A
&$(1.70^{+0.80+0.03+0.50}_{-0.80-0.30-0.00})\times 10^{-2}$  \\
\hline
$B^+ \to D_s^+(K^{*0}(892)\to) K^- \pi^+$
&A
&$(1.40^{+0.80+0.40+0.20}_{-0.30-0.80-0.02})\times 10^{-2}$  \\

$B^+ \to D_s^+(K^{*0}(1410)\to) K^- \pi^+$
&A
&$(0.12^{+0.05+0.01+0.01}_{-0.05-0.03-0.00})\times 10^{-2}$  \\
\hline \hline
\end{tabular}\label{tb2}
\end{center}
\end{table}

\begin{table}[!t]
\caption{The branching ratio (in $10^{-6}$) of the three-body $B/B_s \to DK\pi$ decays with the resonance  $K_2^*(1430)$ in the PQCD approach.}
\begin{center}
\begin{tabular}{l c c}
 \hline \hline
 \multicolumn{1}{c}{Decay Modes}& \multicolumn{1}{c}{Class}&\multicolumn{1}{c}{PQCD }  \\
\hline\hline
$B^0 \to \ D^0(K_2^{*0}(1430)\to)K^+ \pi^-$
&C
&$1.00^{+0.59+0.17+0.06}_{-0.47-0.14-0.05}$    \\

$B_s \to \ D^{0}(\overline{K}_2^{*0}(1430)\to)K^-\pi^+$
&C
&$(3.40^{+2.20+0.50+0.20}_{-1.40-0.50-0.20})\times 10^{-2}$  \\

$B^+ \to \ D^{0}(K_2^{*+}(1430)\to)K^+\pi^0$
&C
&$0.49^{+0.30+0.06+0.02}_{-0.27-0.09-0.02}$     \\

$B_s \to \ D^+(K^{*-}(1430)\to)K^-\pi^0$
&T
&$0.13^{+0.08+0.01+0.01}_{-0.06-0.01-0.01}$      \\

$B_s \to D_s^+(K_2^{*-}(1430)\to)K^+\pi^0$
&T
&$2.59^{+1.81+0.21+0.17}_{-1.07-0.27-0.15}$      \\

$B^0 \to D_s^+(K_2^{*-}(1430)\to)K^-\pi^0$
&E
&$(0.33^{+0.18+0.05+0.02}_{-0.15-0.07-0.04})\times 10^{-2}$    \\

$B^+ \to \ D^+(K_2^{*0}(1430)\to)K^+\pi^-$
&A
&$(0.05^{+0.02+0.01+0.00}_{-0.02-0.01-0.01})$     \\

$B^+ \to D_s^+(K_2^{*0}(1430)\to)K^-\pi^+$
&A
&$(0.39^{+0.20+0.05+0.02}_{-0.20-0.06-0.07})\times 10^{-2}$      \\

 \hline \hline
\end{tabular}\label{tb3}
\end{center}
\end{table}

With the total decay amplitude, the differential branching fraction is written as
\begin{eqnarray}
\frac{d{\cal B} }{dm_{23}}=\tau_B \frac{|\vec p_D||\vec p_K|}{32\pi^3m_B^3}|{\cal A}|^2,
\end{eqnarray}
$\tau_B$ being the $B$ meson lifetime. The magnitudes of three-momenta of one kaon and the bachelor particle in the rest frame of the $K\pi$-pair are given as
\begin{eqnarray}
|\vec p_D|=\frac{\sqrt{\lambda (m_B^2,m_D^2,m_{23}^2})}{2 m_{23}},~~~~~
|\vec p_K|=\frac{\sqrt{\lambda (m_{23}^2,m_K^2,m_{\pi}^2)}}{2 m_{23}},
\end{eqnarray}
with the standard K{\"a}ll{\'e}n function $\lambda(a,b,c)= a^2+b^2+c^2-2(ab+ac+bc)$.

The numerical results of the branching fractions of the $B/B_s\to D K\pi$ decays with the $S$, $P$ and $D$ wave $K\pi$ pairs are summarized in the Tables.~\ref{tb1}, \ref{tb2} and \ref{tb3}, respectively. In the calculations, there are many theoretical uncertainties. Here, we evaluate three main kinds of uncertainties. The first uncertainties origin from the hadronic parameters in the distribution amplitudes of the $B$ mesons, $D$ mesons and $K\pi$ pair, which are nonperturbative but universal. It is noted that they are only calculated from the nonperturbative QCD approach or determined from data. As shown in the tables, this kind of uncertainties are dominate. The second uncertainties are estimated from the unknown higher order and the higher power corrections. Due to the complexity of  calculations, the corrections from higher order and higher power of three-body nonleptonic $B$ decays have not been explored, though part corrections of two-body hadronic $B$ decays have been preformed. In current work, we estimated these uncertainties by choosing $\Lambda_{QCD}=0.25\pm0.05~\rm GeV$ and varying the factorization scales $t$ from $0.8t$ to $1.2t$.  The last uncertainties come from the uncertainties of the CKM matrix element $V_{ub}$. We also note that the experimental measurements of these decays are not available data till now.

In the topological diagram approach, the amplitude can be decomposed in term of graphical contributions. The relevant graphs consist of the following (1) a (color-favored)``tree" amplitude T, associated with the transition $\bar b \to \bar u q \bar c$  ($q = d~{\rm or} ~s$) in which the $\bar c q$ system forms a color-singlet $D^+$ or $D_s^+$ meson; (2) a ``color-suppressed" amplitude C, associated with the transition $\bar b \to \bar c u \bar q$ in which the $\bar c u$ system forms a color-singlet $D^0$ meson; (3) an ``exchange" amplitude E in which the $\bar b$ quark and an initial $q$ quark in the decaying neutral $B$ meson exchange a $W$ and become $\bar c$ and $u$; and (4) an ``annihilation" amplitude A contributing only to charged $B$ decay through the subprocess $\bar bu \to  \bar q c$ by means of a $W$ in the direct channel. In the tables, all the decays are thus classified according to the dominant contribution.

In order to include the effects of resonance, the time-like form factor $F_S(m_{ij}^2)$ is introduced in the distribution amplitudes of two-meson wave functions \cite{Wang:2015uea}. In particular, this form factor is  parameterized by the relativistic Breit-Wigner (RBW) model \cite{Breit:1936zzb}, which has been adopted extensively in experimental analysis and been regarded as a valid model for describing a narrow resonances, such as resonances $K^*(892)$ and the $K_2^*(1430)$. However, RBW model fails to describe the $S$-wave resonance $K_0^*(1430)$,  because the resonance interferes strongly with a slowly varying non-resonant term. To help resolve this dilemma, the so-called LASS model is developed,  which consists of the resonance as well as an effective-range nonresonant component. Readers are referred to the references \cite{Aston:1987ir, Back:2017zqt, 112} for details. Therefore, for these S-wave decays we list three types of branching fractions, i.e., the total branching fractions (LASS) including both resonant and non-resonant contributions, the branching fractions including only the resonant contributions and the branching fractions (LASSNR) including contributions from non-resonance, as shown in Table.~\ref{tb1}. An alternative scenario has also been proposed \cite{Wang:2020saq}, where the form factor $F_S(m_{ij}^2)$ was derived from the matrix element of the vacuum to $K\pi$ final state and was related to the corresponding scalar time-like form factor $F_0^{K\pi}(m_{ij}^2)$.

Compared with the $B/B_s\to \overline{D}K\pi$ decays calculated in the ref.\cite{Zou:2022xrr}, $B/B_s\to DK\pi$ decays are suppressed by the CKM matrix elements $\mid V_{ub}/V_{cb}\mid^2$, especially for those strangeness decays. Thus, the branching fractions of the $B/B_s\to DK\pi$ decays with the vector resonances $K^*(892)$ and $K^*(1410)$ are smaller than those of corresponding $B/B_s\to \overline{D}K\pi$ decays. From the Table.~\ref{tb2}, one finds that the branching fraction of the $B_s\to D_s^+(K^{*-}(892)\to)K^-\pi^0$ is much larger than those of other decay modes, and it is at the order of $10^{-5}$. In fact, this decay is a ``T'' type channel, which is enhanced by both the large WC $C_1/3+C_2$ and the large CKM matrix element $V_{cs}$. However, due to the suppression by the small CKM element $V_{cd}$, another ``T'' type decay channel $B_s\to D^+(K^{*-}(892)\to)K^-\pi^0$ has a small branching fraction, the order of which is equivalent to those of color-suppressed channels with large CKM element $V_{cs}$, such as the $B^+\to D^0(K^{*+}(892)\to)K^+\pi^0$ decay. In calculating the decays with a light meson emitted, two nonfactorizable diagrams are cancelled by each other significantly, then their total amplitude is smaller than that of factorizable diagrams. However, for the color-suppressed decay modes with a $D$ meson emitted, because the mass of charm quark is much larger than that of light quark, the cancellation between two nonfactorizable diagrams does not exist any more, which leads that the sum of two amplitudes of two nonfactorizable diagrams becomes sizable. Therefore, these two nonfactorizable diagrams with the large WC $C_2$ dominate the decay amplitude. We thus suggest that the experimentalists can first observe those CKM enhanced ``T" and ``C" type decays with the branching ratios at the order of $10^{-6}$ or even bigger.

We note that for the $S$ and $D$ waves the branching fractions of $B/B_s\to D (K_{0(2)}^+(1430)\to)K\pi$ decays are close to or even larger than those of $B/B_s\to \overline{D} (K_{0(2)}^+(1430)\to) K\pi$ decays, for example,  $\mathcal{B}(B^0\to D_s^+(K_{0}^-(1430)\to) K^+ \pi^0)\sim7.1\times10^{-6} > \mathcal{B}(B^0 \to D^-(K_{0}^+(1430)\to)K^+\pi^0)\sim 1.2\times10^{-6}$. In fact, the decays $B/B_s\to D (K_{0(2)}^+(1430)\to)K\pi$ with $D$ meson emission are color suppressed, while the decays $B/B_s\to {\overline D} (K_{0(2)}^+(1430)\to)K\pi$ are color-favored with $K\pi$-pair emission. When the $K\pi$-pair is a $S$-wave, the factorizable emission diagrams with large WCs are highly suppressed by the tiny vector decay constant of the emitted scalar structure. Similarly, if the emitted $K\pi$-pair is a $D$-wave, the contributions from the factorizable emission diagrams vanish, bencause the tensor structure can not be produced through the $V-A$ currents. As stated above, the two nonfactorizable emission diagrams with small WC $C_1$ are also cancelled by each other. For $B^0\to D_s^+(K_{0,2}^-(1430)\to)K^+\pi^0$ decays, $D_s^+$ meson is emitted and the factorizable emission diagrams with large WCs $C_1/3+C_2$ dominate the whole amplitudes, leading to large branching fractions.

{
\begin{table}[!t]
\caption{The branching ratios (in $10^{-6}$) of $B_{(s)}\to D K^*$ decays probed from the quasi-two-body $B_{(s)}\to D (K^*\to) K\pi$ decays based on the narrow-width-approximation(NWA),  together with the experimental data \cite{ParticleDataGroup:2022pth} and the former PQCD predictions from refs.\cite{Zou:2009zza,Zou:2016yhb,Zou:2012sx}.}
\begin{center}
\begin{tabular}{l c c c}
 \hline \hline
 \multicolumn{1}{c}{Decay Modes}&\multicolumn{1}{c}{NWA} &\multicolumn{1}{c}{FormerPQCD} \\
\hline\hline
$B^0 \to \ D^0K^{*0}(892)$
&$2.94^{+1.69}_{-1.45}$  &$1.92^{+1.22}_{-0.94}$\\

$B^0 \to \ D^0K^{*0}(1410)$
&$4.77^{+2.50}_{-2.18}$  &....\\

$B^0 \to \ D^0K_0^{*0}(1430)$
&$1.95^{+1.03}_{-0.96}$ & $1.02^{+0.88}_{-0.71}$ \\

$B^0 \to \ D^0K_2^{*0}(1430)$
&$3.00^{+1.85}_{-1.48}$   &$4.18^{+1.98}_{-1.66}$   \\

\hline
$B_s \to \ D^{0}\overline{K}^{*0}(892)$
&$0.12^{+0.08}_{-0.05}$  &....\\

$B_s \to \ D^{0}\overline{K}^{*0}(1410)$
&($0.19^{+0.13}_{-0.07})\times 10^{-2}$  &....\\

$B_s \to \ D^{0}\overline{K}_0^{*0}(1430)$
&$0.07^{+0.05}_{-0.02}$ & $0.06^{+0.05}_{-0.04}$  \\

$B_s \to \ D^{0}\overline{K}_2^{*0}(1430)$
&$0.10^{+0.07}_{-0.05}$   &$0.14^{+0.08}_{-0.06}$   \\

\hline
$B_s \to \ D^+K^{*-}(892)$
&$1.8^{+0.92}_{-0.45}$  &$1.42^{+0.65}_{-0.51}$\\

$B_s \to \ D^+K^{*-}(1410)$
&$3.04^{+1.38}_{-0.94}$  &....\\

$B_s \to \ D^+K^{*-}(1430)$
&$0.95^{+0.54}_{-0.42}$ & $0.76^{+0.81}_{-0.71}$ \\

$B_s \to \ D^+K^{*-}(1430)$
&$0.75^{+0.24}_{-0.18}$   &$1.12^{+0.58}_{-0.48}$   \\

\hline

$B^+ \to \ D^{0}K^{*+}(892)$
&$3.00^{+1.42}_{-1.66}$  &$2.05^{+1.46}_{-0.92}$\\

$B^+ \to \ D^{0}K^{*+}(1410)$
&$4.91^{+2.80}_{-1.26}$  &....\\

$B^+ \to \ D^{0}K_0^{*+}(1430)$
&$1.25^{+0.92}_{-0.88}$ & $2.13^{+1.42}_{-1.33}$ \\

$B^+ \to \ D^{0}K_2^{*+}(1430)$
&$2.91^{+1.82}_{-1.44}$   &$3.73^{+1.66}_{-1.56}$  \\

 \hline

$B_s \to D_s^+K^{*-}(892)$
&$40.00^{+20.70}_{-9.60}$  &$33.10^{+15.70}_{-12.28}$\\

$B_s \to D_s^+K^{*-}(1410)$
&$68.40^{+35.52}_{-15.96}$  &....\\

$B_s \to D_s^+K_0^{*-}(1430)$
&$22.90^{+12.06}_{-5.38}$ & $14.50^{+8.08}_{-6.31}$  \\

$B_s \to D_s^+K_2^{*-}(1430)$
&$15.60^{+10.98}_{-6.68}$   &$20.60^{+11.90}_{-10.10}$   \\

\hline
$B^0 \to D_s^+K^{*-}(892)$
&$(1.6^{+0.8}_{-0.8})\times 10^{-2}$  &$(1.68^{+0.77}_{-0.55})\times 10^{-2}$\\

$B^0 \to D_s^+K^{*-}(1410)$
&$(2.30^{+1.40}_{-1.60})\times 10^{-2}$  &....\\

$B^0 \to D_s^+K_0^{*-}(1430)$
&$(4.50^{+1.60}_{-1.00})\times 10^{-2}$ & $....$  \\

$B^0 \to D_s^+K_2^{*-}(1430)$
&$0.02^{+0.01}_{-0.01}$   &$0.06^{+0.02}_{-0.02}$    \\

\hline
$B^+ \to \ D^+K^{*0}(892)$
&$0.31^{+0.16}_{-0.09}$  &$0.11^{+0.03}_{-0.05}$\\

$B^+ \to \ D^+K^{*0}(1410)$
&$0.38^{+0.21}_{-0.19}$  &....\\

$B^+ \to \ D^+K_0^{*0}(1430)$
&$0.15^{+0.15}_{-0.06}$ & $0.41^{+0.19}_{-0.20}$  \\

$B^+ \to \ D^+K_2^{*0}(1430)$
&$0.15^{+0.06}_{-0.07}$   &$0.527^{+0.20}_{-0.19}$   \\

\hline

$B^+ \to D_s^+K^{*0}(892)$
&$(2.10^{+1.30}_{-1.30})\times 10^{-2}$  &$(0.50^{+0.21}_{-0.21})\times 10^{-2}$\\

$B^+ \to D_s^+K^{*0}(1410)$
&$(2.70^{+1.10}_{-1.30})\times 10^{-2}$  &....\\

$B^+ \to D_s^+K_0^{*0}(1430)$
&$(1.20^{+0.90}_{-0.70})\times 10^{-2}$ & $(2.50^{+1.00}_{-1.30})\times 10^{-2}$  \\

$B^+ \to D_s^+K_2^{*0}(1430)$
&$(1.20^{+0.60}_{-0.70})\times 10^{-2}$   &$(3.40^{+1.40}_{-1.30})\times 10^{-2}$  \\

 \hline \hline

\end{tabular}\label{tb4}
\end{center}
\end{table}
}

Under the narrow-width-approximation, a branching fraction of the quasi-two-body decay can be decomposed as
\begin{eqnarray}
\mathcal{B}[B\to M_1R\to M_1M_2M_3]=\mathcal{B}[B\to M_1R]\times\mathcal{B}[R\to M_2M_3],
\end{eqnarray}
$R$ being a resonance. This approximation provides us an effective approach to extract the branching fractions of the $B\to D K^*$ decays, by combining the calculated branching fractions of the above quasi-two-body decays and the experimental data of the $K^*\to K\pi$ decays. In turn, the comparison between extracted branching fraction and the available experimental data is also helpful to test our theoretical predictions.

In the experimental side, the branching fractions of the $K^*\to K\pi$ have been measured with high precision  \cite{ParticleDataGroup:2022pth}, and the results are given as
\begin{eqnarray}
\mathcal{B}(K^*(892)\to K\pi)& \thicksim & 1,\\
\mathcal{B}(K^*(1410)\to K\pi)&=&(6.6\pm1.2)\%,\\
\mathcal{B}(K^*_0(1430)\to K\pi)&=&(93\pm10)\%,\\
\mathcal{B}(K_2^*(1430)\to K\pi)&=&(49.9\pm1.2)\%.
\end{eqnarray}
Within above data and the numerical results listed in tables, we can estimate the branching fractions of the $B/B_s\to D K^*$ decays, and the results are presented in Table.~\ref{tb4}. It is found that the new obtained branching fractions are consistent to the theoretical results based on PQCD approach within the uncertainties.


It is well known to us that the direct $CP$ asymmetries of $B$ meson decays origin from the interference between the tree and the penguin contributions. However, in SM only tree operators contribute to the considered quasi-two-body $B/B_s\to D (K^*\to)K\pi$ decays, which means that there are no direct $CP$ asymmetries in these decays. If a large $CP$ asymmetry were measured in future, it would be a direct signal of new physics beyond SM.

Currently, the theoretical study of three-body $B$ decays is still challenging, because it is hard for us to separate the resonant and non-resonant contributions, or perturbative and nonperturbative part. Therefore, the theoretical description of the three-body decays is still in the stage of the modeling now \cite{Virto:2016fbw}. In this work, we only discuss the quasi-two-body $B$ decays where two final particles move collinearly and the bachelor recoils back. In this case, the factorization hypothesis might be valid, and the amplitude can be factorized into three parts, namely the wilson coefficient, the hard kernel and the nonperturbative wave functions. In the calculation, since the wave functions of light mesons and initial $B$ mesons have been well studied in two-body $B$ decays \cite{Bell:2013tfa,Wang:2019msf}, the most important inputs are the wave functions of the two-meson pair, which cannot be obtained from the first principle till now. Here, the wave functions of $K\pi$-pair for different waves are from the phenomenological models, and the inner parameters are determined from the experimental data. Although two meson wave functions have been investigated based on light-cone QCD sum rules recently \cite{Hambrock:2015aor,Descotes-Genon:2023ukb}, and we still call for more results with high precision. Once the $K\pi$-pair wave functions were well studied, the uncertainties of theoretical predictions would be reduced remarkably, which will be helpful for us to study the three-body $B$ decays on a deeper level further.
\section{Summary} \label{sect:4}
Within PQCD approach, we calculate the branching fractions of the quasi-two-body $B_{(s)}\to D (R\to K\pi)$ decays, $R$ being the vector resonance $K^*(892)$, the scalar resonance $K_0^*(1430)$ and the tensor one $K_2^*(1430)$. In the calculations, we adopted the wave functions of the $K\pi$ pair that are determined in previous studies of the $B \to \overline{D}(K^*\to) K\pi$ decays. Due to the suppression of the small CKM elements $V_{ub}V_{cd(s)}$, especially for those decays with the four-quark operators without strange quark, the resonant branching fractions are in range of $10^{-9}\sim10^{-6}$. The T-type $B_s\to D_s^+(K\pi)^-$ decays enhanced by $V_{cs}$ have large branching fractions which would be measurable in the on-going LHCb experiment. Other results could also be tested in LHCb or Belle-II experiments. Furthermore, in order to further verify the reliability of the $K\pi$ pair wave function, we extracted the branching fractions of the corresponding two-body $B/B_s\to DK^*$ decays using the narrow-width-approximation, and the results agree well with the previous theoretical predictions. There are no local direct $CP$ asymmetries because all decays are governed by only tree operators in SM.
\section*{Acknowledgment}
This work is supported in part by the National Science Foundation of China under the Grant No. 11705159 and No. 12375089, and the Natural Science Foundation of Shandong province under the Grant No. ZR2022MA035, no. ZR2019JQ04 and No. ZR2022ZD26.

\bibliographystyle{bibstyle}
\bibliography{refs}
\end{document}